# Initializing, manipulating and storing quantum information with bismuth dopants in silicon

Gavin W. Morley[1,2*], Marc Warner[1,2], A. Marshall Stoneham[1,2], P. Thornton Greenland[1,2], Johan van Tol[4], Christopher W. M. Kay[1,3] & Gabriel Aeppli[1,2]

[1] *London Centre for Nanotechnology, University College London, London WC1H 0AH, UK*

[2] *Department of Physics and Astronomy, University College London, London WC1E 6BT, UK*

[3] *Institute of Structural and Molecular Biology, University College London, London WC1E 6BT, UK*

[4] *National High Magnetic Field Laboratory and Florida State University, Tallahassee, Florida 32310, USA*

*e-mail: g.morley@ucl.ac.uk

**A prerequisite for exploiting spins for quantum data storage and processing is long spin coherence times. Phosphorus dopants in silicon (Si:P) have been favoured[1,2,3,4,5,6,7,8] as hosts for such spins because of measured electron spin coherence times ($T_2$) longer than any other electron spin in the solid state: 14 ms at 7 K[9]. Heavier impurities such as bismuth in**

**silicon (Si:Bi) could be used in conjunction with Si:P for quantum information proposals that require two separately addressable spin species[10,11,12,13]. However, the question of whether the incorporation of the much less soluble Bi into Si leads to defect species that destroy coherence has not been addressed. Here we show that schemes involving Si:Bi are indeed feasible as the electron spin coherence time $T_2$ exceeds 1 ms at 10 K. We polarized the Si:Bi electrons and hyperpolarized the I=9/2 nuclear spin of $^{209}$Bi, manipulating both with pulsed magnetic resonance. The larger nuclear spin means that a Si:Bi dopant provides a 20-dimensional Hilbert space rather than the four dimensional Hilbert space of an I=1/2 Si:P dopant.**

Kane's suggestion for a Si:P quantum computer[1], where the electron and nuclear spins of P impurities are regulated and read out using electrical gates, has inspired many researchers. Two particular challenges in building the Kane quantum computer are placing phosphorus dopants with atomic precision[2] below the surface, and depositing metallic contacts between them.

Alternate schemes[10,11,12,13], that are conceptually more complex but impose less stringent requirements on fabrication, take advantage of other group V elements which also substitute for silicon. With qualitatively similar behaviour to Si:P dopants, these other dopants exhibit electron spin resonance (ESR) at field/frequency combinations distinct[14] from Si:P, allowing selective excitation and detection with microwave pulses. Because of their lower solubility and higher binding energies, they have been of far less relevance to microelectronics, and so in contrast to Si:P, their spin relaxation has remained relatively unexplored. This is particularly true for the heaviest element, bismuth, which has the highest binding energy and also the largest nuclear spin (I=9/2), both of which should be advantageous for quantum computing because they would permit higher temperature operation and a larger auxiliary state space for quantum data storage, respectively. While Bi looks attractive in principle, a potential problem is that Bi is the largest and least soluble of the group V elements[15] – if the site of incorporation is too distorted, additional sources of decoherence could be active. Accordingly, we have used pulsed ESR to measure the spin-lattice relaxation time $T_1$ and the decoherence time $T_2$, as well as demonstrating the controlled preparation of quantum states (via Rabi oscillations) and nuclear spin manipulations via pulsed electron-nuclear double resonance (ENDOR).

Figure 1 depicts the ESR spectra obtained with both standard (9.7 GHz) and very high (240 GHz) frequency microwave radiation. In each case Si:Bi yields the ten resonances expected for an electron (spin ½) coupled to a nuclear spin of $^9/_2$. The resonant fields of all transitions are well simulated as shown in the figure and described in the Supplementary Information. The Gaussian linewidth is shown in the Supplementary Information to be 0.41 ± 0.002 mT with 9.7 GHz radiation which agrees with the value reported previously[14] and attributed to interactions with the natural (4.7%) concentration of $^{29}$Si nuclear spins.

Figure 1a shows a spin echo-detected field-swept spectrum recorded with a pulsed ESR spectrometer (Bruker E580) operating at 9.7 GHz. The Supplementary Information describes in more detail all pulse sequences used in our experiments. The spectrum in Figure 1b was recorded in continuous-wave (CW) mode at 240 GHz with a quasi-optic spectrometer[16,17] at the National High Magnetic Field Laboratory in Tallahassee, Florida.

For the 240 GHz experiments, a magnetic field above 8.3 T together with a temperature of 3 K ensures that the electron spin polarization is above 95%, providing a good initial state for a quantum computation using the electron spin. This almost pure state avoids the problems encountered by liquid state NMR quantum computers with the use of pseudo-pure starting states[18].

Additionally, we have initialized the electron-nuclear spin system by transferring some of the large electron polarization to the bismuth nuclei. Figure 1c shows that this was achieved with above-band-gap white light without the need for the resonant excitation used in some previous experiments[19]. This dynamic nuclear polarization (DNP) is due to the Overhauser effect[20] whereby the electron spin of the photoelectrons relaxes by 'flip-flopping' with the $^{209}$Bi nuclear spin. The photoelectrons are initially unpolarized and to move towards thermal equilibrium it is necessary for ~45% of them to 'flip' spins. These flips can conserve angular momentum if a $^{209}$Bi nuclear spin 'flops' in the other direction. The energy required to flop a nuclear spin is negligible compared to the thermal energy in our experiment. Similar effects have been seen with

$^{29}$Si nuclear spins in silicon[21] and with Si:P[6,7], while the application of light that is only slightly larger than the band gap favours the formation of bound excitons in Si:Bi and a different mechanism for DNP[22]. The magnitude of the $^{209}$Bi nuclear polarization near the surface may be much larger than in the bulk because the light does not penetrate deeply into the sample[6].

To use the full twenty-dimensional Hilbert space available from a Si:Bi donor for quantum computing, it is necessary not only to initialize the spin system and manipulate the electron spin, but also to manipulate the nuclear spin. To demonstrate the feasibility of this we performed pulsed electron-nuclear double resonance (ENDOR)[23] at 240 GHz as illustrated in Figure 2a. We fit the ENDOR resonance with a 0.24 MHz Gaussian line but this linewidth is not due to the nuclear relaxation times of the $^{209}$Bi. Some strain in the crystal may broaden the line via the nuclear quadrupole interaction[23].

To characterize the quality of our electron spin manipulations we recorded Rabi oscillations as shown Figure 2b, obtaining a spin flip time of 13 ns. The characteristic timescale for the decay of these oscillations is around 100 ns, but this only provides a lower bound on the time for the decay of spin coherence. The 0.41 mT CW linewidth means that not all of the spins in our sample were on resonance and this may dominate the Rabi decay we observe.

To measure the spin coherence times of Si:Bi we recorded the electron spin echo size as a function of the separation ($\tau/2$) of the $\pi$ refocusing pulse and the initial $\pi/2$ pulse (the inset of Figure 3a shows the pulse sequence). For temperatures above ~20 K the decay is exponential: $e^{-\tau/T_2}$. The exponential decay constant for spins in a solid, $T_m$, is referred to here as $T_2$ because the electron spin density is low enough that these spins interact only weakly. At lower temperatures the coherence decay is clearly non-exponential as shown in Figure 3a. We fitted this decay with the same function[24] that has been used for similar experiments with Si:P: $e^{-\tau/T_2 - \tau^3/T_S^3}$. $T_S$ characterizes the coherence decay due to the presence of $^{29}$Si nuclear spins with natural abundance of 4.6%. Figure 3a shows that the $T_S$ decay destroys the Hahn echo signal after about $\tau = 0.6$ ms, making it hard to measure $T_2$ times longer than this.

Figure 4 shows the temperature ($T$)-dependent $T_2$ and $T_S$ times. For $T>20$ K the spin decoherence is dominated by the spin-lattice relaxation time, $T_1$, which we measured with the inversion recovery pulse sequence[23] (Figure 3b). The $T_1$ time characterizes the time taken for the electron spins to polarize. It can be thought of as the timescale for storing classical information, in contrast to the quantum information storage time, $T_2$. As expected, the $T_1$ is dominated by phonons, but most of these have energies that are much larger than the energy gap between spin up and spin down. As a result, two phonon processes (such as the absorption of a high-energy phonon and the emission of an even higher energy phonon) occur more frequently than single-phonon processes. The dependence of the spin lattice relaxation rate, $1/T_1$, on $T$ is well described by an $e^{-\Delta E/T}$ Orbach term (two phonon process via an excited state) added to a $T^7$ Raman term (two phonon process using a virtual excited state) and a single-phonon term proportional to $T$. The value for $\Delta E$ used in Figure 4 was 500 K, but values from 450-800 K provide an acceptable fit. A value of $\Delta E = 400$ K has previously been measured with ESR of a more concentrated sample, and identified[25] with the energy gap to the 1s($T_2$) orbital excited state.

Si:Bi has a larger ionization energy than the other group V donors, which reduces the number of phonons with enough energy to access the excited $1s(T_2)$ states where two-phonon Orbach spin-lattice relaxation occurs strongly[25]. This means that in the temperature regime where Orbach effects dominate, the $T_1$ time of Bi is longer than all other group V donors in Si. The $T_1$ times measured here follow the same temperature dependence as those in Ref. [26], but the higher spin concentration of $4 \times 10^{16}$ cm$^{-3}$ used in that work lead to shorter relaxation times. A similar concentration-dependent effect[27] has been described for Si:P.

For temperatures below 20 K, the spin echo decay in Si:Bi is limited by the 4.6% of Si nuclei with spin ½ which provide the $T_S$ decay. With an isotopically pure sample of $^{28}$Si:Bi this contribution would not have been present, revealing the magnitude of other interactions such as those with nearby Si:Bi electron spins.

The largest room-temperature electron spin coherence time[29] reported is 1.8 ms, but this was in (isotopically pure) diamond which has not yet been integrated with commercial semiconductor fabrication. Removing the unwanted neighbouring spins could increase the Si:Bi $T_2$ towards the limiting value of $2T_1$.

The diamond results were obtained with a single electron spin which is closer to a quantum computation than our experiments with >$10^{12}$ electron spins. Electrical detection may permit the readout of single electron spins in silicon[4] and the differing resonant frequencies of dopants in slightly different environments may permit selective addressing[11].

The long spin coherence times we measure show that Si:Bi is well suited for storing quantum information. We have flipped the electron spin in a time of 13 ns, and find $T_2$ (>1 ms) $10^5$ times longer. We conclude that quantum information processing in silicon can be based not only on phosphorus dopants, but also bismuth dopants and combinations of the two.

**Summary of Methods**

The samples used here are single float-zone crystals of silicon, bulk doped in the melt with bismuth atoms. A sample with concentration estimated as $3 \times 10^{15}$ Bi cm$^{-3}$ from Fourier-transform infra-red spectroscopy[30] was used for the 9.7 GHz measurements, while a ~$4 \times 10^{14}$ Bi cm$^{-3}$ sample was used for the measurements at 240 GHz. The white light applied for Figures 1c and 2a was generated by a continuous filtered Xe lamp. For the pulsed ENDOR measurement (Figure 2a) we applied white light to shorten the electron spin $T_1$ time[27], enabling a shorter shot repetition time.

The highest field ESR resonance was used for all of the relaxation times presented here. To measure the Hahn echo decays below 20 K(such as Figure 3a), single shots were collected and the magnitude of the echo signals were averaged[9,16].

**Acknowledgements** We thank Bernard Pajot for supplying the silicon samples. Our research was supported by RCUK through the Basic Technologies program and a Wolfson Royal Society Research Merit Award. The National High Magnetic Field

Laboratory is supported by NSF Cooperative Agreement No. DMR-0654118, and by the State of Florida. GWM is supported by an 1851 Research Fellowship.

**Author Contributions** G.W.M., M.W., and C.W.M.K carried out the experiments at 9.7 GHz, G.W.M. and J.v.T. carried out the experiments at 240 GHz. G.W.M. and P.T.G. performed the simulations. G.W.M. analyzed the data which were interpreted by G.W.M., A.M.S., J.v.T., C.W.M.K. and G.A.. The paper was written by G.W.M., A.M.S., J.v.T., C.W.M.K. and G.A..

**Additional Information** Supplementary information accompanies this paper on www.nature.com/naturematerials. Reprints and permissions information is available online at http://npg.nature.com/reprintsandpermissions. Correspondence and requests for materials should be addressed to G.W.M.

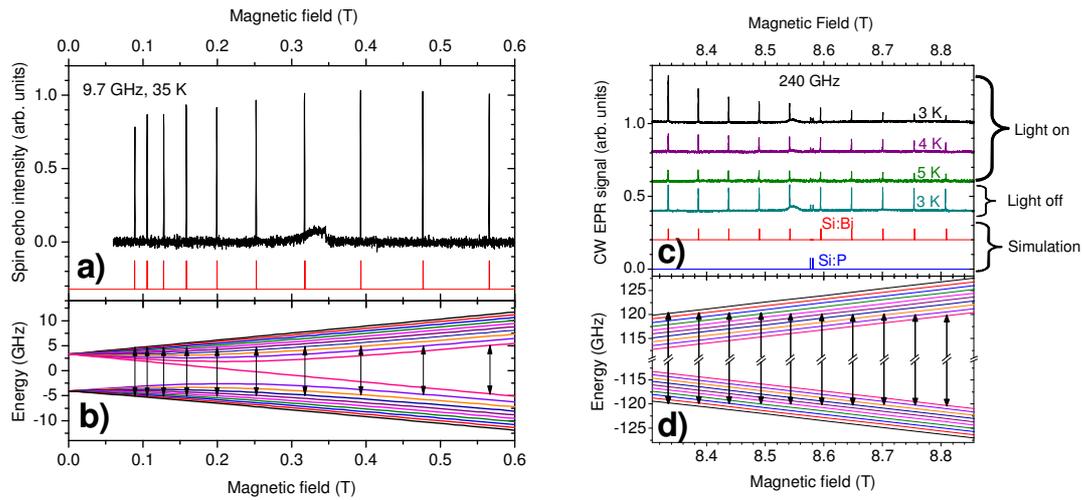

Figure 1. EPR spectra of Si:Bi with a) a frequency of 9.7 GHz and c) a frequency of 240 GHz. At the high magnetic field required for the higher frequency experiment, the ten Si:Bi resonances are evenly spaced (c). This is not the case for the lower frequency experiment as the correspondingly lower magnetic field is not strong enough to define the axis of quantization: the large hyperfine interaction of a = 1.4754 GHz (for definition see Supplementary Information) is comparable in size to the Zeeman term in the Hamiltonian. The red lines are simulations of the Si:Bi resonances and the blue line is a simulation of the Si:P resonances which account for the sharp but weak features visible in the high field data. In all spectra we attribute the small broad signal around *g* = 2 (0.34 T for 9.7 GHz and 8.55 T for 240 GHz) to dangling bond defects in the silicon. $^{209}$Bi nuclear polarization manifests itself in the form of a larger signal for the low-field resonance lines at temperatures below 5 K when above-band-gap light is applied. The spectra have been offset for clarity and the arbitrary units are the same for each spectrum. b) and d) show the simulated energy levels as a function of low and high magnetic field

respectively, with the arrows indicating transitions that flip the electron spin state but leave the nuclear spin state unchanged.

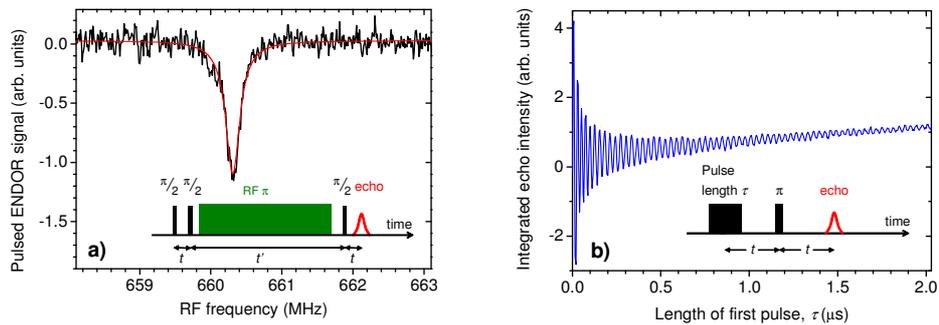

Figure 2. Manipulating the nuclear and electronic spins of Si:Bi a) Pulsed ENDOR manipulates the Bi nuclear spin as well as the electron spin. A microwave frequency of 240 GHz (black rectangles in inset which control the electron spin) was used at a temperature of 3 K and the length of the RF pulse (green rectangle that controls the nuclear spin) was 150 µs. b) Rabi oscillations of the electron spin at 25 K with 9.7 GHz radiation. The spin is flipped using a pulse of $\tau$ = 13 ns duration.

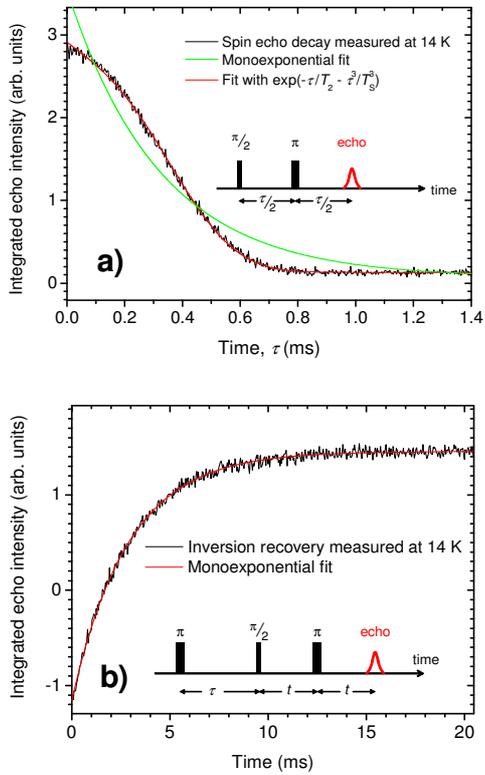

Figure 3. Spin relaxation of Si:Bi at 9.7 GHz. a) Spin echo decay measured at 14 K. b) Inversion recovery measurement of $T_1$ at 14 K.

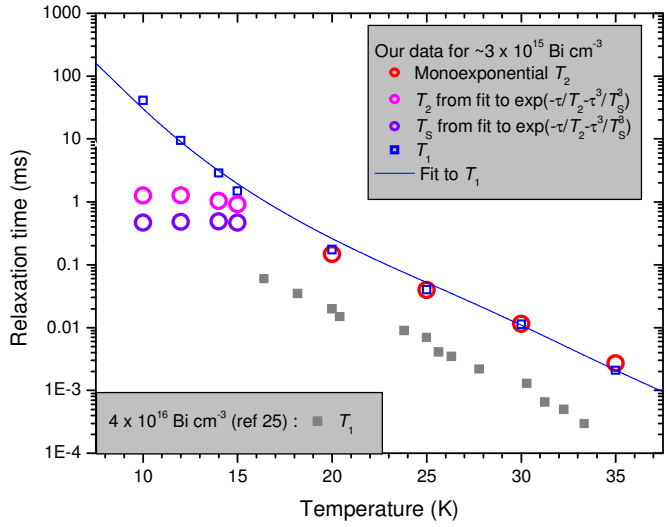

Figure 4. Spin relaxation times as a function of temperature for Si:Bi and Si:P. The open shapes are data measured here for a sample with concentration ~ 3 × $10^{15}$ Bi cm$^{-3}$. Blue squares are the $T_1$, red circles are the monoexponential $T_2$ values, pink and purple circles are the $T_2$ and $T_S$ values respectively from the fit to the function $e^{-\tau/T_2 - \tau^3/T_S^3}$. The blue line is a fit to our $T_1$ measurements described in the main text. For comparison, the grey solid squares are $T_1$ measurements by Castner[26] for a more concentrated (4 × $10^{16}$ cm$^{-3}$) sample of Si:Bi.

## Supplementary Information

## Simulation

The resonant magnetic fields of all of the resonances in Figure 1 are well simulated by the isotropic spin Hamiltonian

$$H = g\,\mu_B\,\mathbf{B} \bullet \mathbf{S} - g_N\,\mu_N\,\mathbf{B} \bullet \mathbf{I} + a\,\mathbf{S} \bullet \mathbf{I} \qquad (1)$$

using the published values[14] for the hyperfine coupling ($a$ = 1.4754 GHz) and the electronic $g$-factor ($g$ = 2.0003) as well as the accepted nuclear $g$-factor of $g_N$ = +0.914. The Bohr and nuclear magnetons are $\mu_B$ and $\mu_N$ respectively; $\mathbf{S}$ and $\mathbf{I}$ are the electronic and nuclear spin operators respectively and $\mathbf{B}$ is the applied magnetic field. The first two terms in the spin Hamiltonian describe the electronic and nuclear Zeeman couplings respectively and the last term is the hyperfine coupling between electronic and nuclear spins. We diagonalised equation (1) using Mathematica software. The microwave frequencies used in Figure 1a and Figure 1b were 9.67849 GHz and 240.00000 GHz respectively. Although Figure 1b is a field modulated CW spectrum the resonances have a Gaussian lineshape rather than a first-derivative Gaussian lineshape because the long spin-lattice relaxation time leads to saturation[14,19]: the timescale for exciting the ESR transitions is comparable with (or faster than) the timescale for $T_1$ spin relaxation. The resonances are larger in the dark because the electronic $T_1$ is longer, exaggerating the saturation effect.

At low temperatures, an electron is bound to the [209]Bi dopant and a spectrum similar to those in Figure 1 can be observed. At room temperature many of the bismuth atoms are ionized, giving up their electrons to the conduction bands, so that these dopants act as donors: computer chips use phosphorus donors, which are more easily ionized, to supply mobile electrons.

We have considered the lowest twenty energy levels in Si:Bi here. Excited states also exist which can be excited with infra-red radiation[5]. Note that the similar terminologies used for the excited state $1s(T_2)$ and the $T_2$ time is purely coincidental.

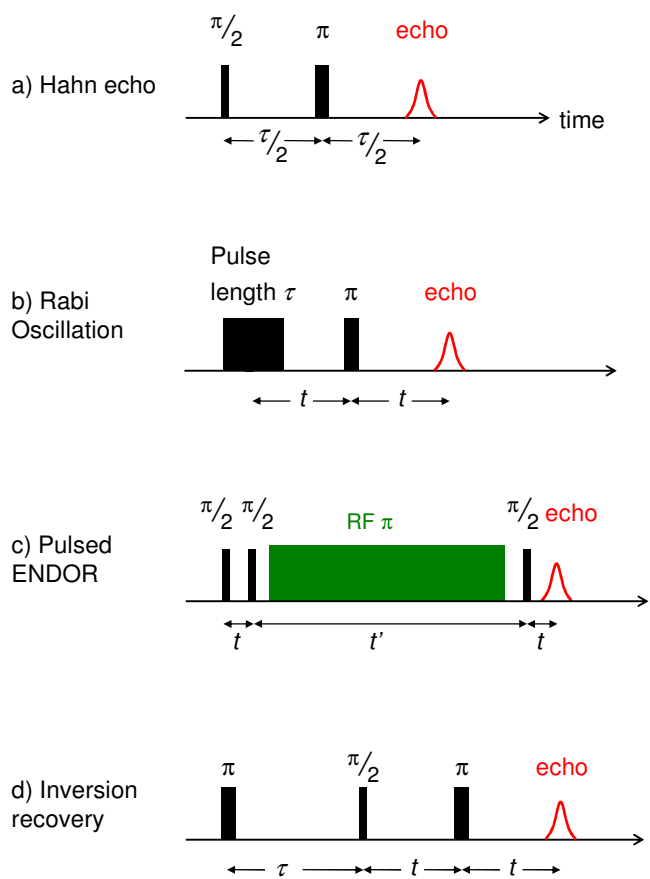

Figure S1. Magnetic resonance pulse sequences described in reference 23. a) Hahn echo. Plotting the integrated echo intensity against the applied magnetic field, while keeping $\tau$ constant provides a spectrum showing the electron spin resonances present with spin dephasing times longer than $\sim \tau$. Figure 1a shows an example of this echo-detected field sweep. When the field is set to a resonance, plotting the integrated echo intensity against $\tau$ generates a decay that is characteristic of spin dephasing in the sample. For non-interacting spins the decay will be exponential with time constant $T_2$. To measure spin dephasing times longer than 0.1 ms (i.e. for temperatures below 20 K) we used 'magnitude detection' as described in reference 9 to reduce the phase noise from the pulsed ESR spectrometer: instead of averaging the in-phase and out-of-phase signals separately, we added the in-phase and out-of-phase signals in quadrature before averaging these magnitudes. To fit these decays some researchers[8, supplementary ref 1] have used the function $e^{-\tau/T_2 - \tau^n/T_S^n}$ allowing $n$ to be a free parameter, rather than taking $n=3$ as used by Chiba and Hirai[24]. This produces an equally good fit to our data, with longer values of $T_2$ and best-fit values of $n$ between 2.5 and 3 depending on the temperature. b) Rabi oscillation sequence with Hahn echo detection. The length of the first pulse is initially negligible and is then increased incrementally. The magnetic field was set to the resonance at 0.253 Tesla. c) Mims ENDOR uses a stimulated electron spin echo composed of three electron spin π/2 pulses that were 800 ns long in our experiment with 240 GHz frequency. The nuclear pulse was 150 μs long and the frequency of this was swept through the nuclear resonance transition. Figure 2b shows the lowest frequency ENDOR resonance from electron spin -1/2, nuclear spin +9/2 to electron spin -1/2, nuclear spin +7/2. d) The inversion recovery sequence measures $T_1$ by incrementally increasing $\tau$ for fixed $t$.

**Linewidth of resonances**

Figure S2 shows expanded views of the high-field resonance lines observed with 9.7 GHz and 240 GHz. They are well fit by Gaussian lineshapes having widths of 0.41 ± 0.002 mT and 0.50 ± 0.004 mT respectively. The latter value comes from a saturated CW measurement so may be larger than an unsaturated measurement.

A previous report[14] of the ESR linewidth of Si:Bi at ~9 GHz found 0.45 ± 0.02 mT but estimated that this saturated CW measurement overestimated the linewidth by

~8%. The agreement is therefore good with our measurement at 9.7 GHz which does not suffer from saturation problems.

The linewidth of Si:P is known to be due to the nuclear spins of the naturally abundant 4.7% $^{29}$Si present[14]. As a result of quantum interference between the valleys in the Si bandstructure, a few Si sites close to the phosphorus donor provide most of the linewidth[14]. The Si:Bi electronic wavefunction is smaller than that of Si:P enhancing the strength of the interaction between the Si:Bi electron and $^{29}$Si nuclei on the important sites. This explains the observation that the Si:Bi ESR linewidth is somewhat larger than that of Si:P.

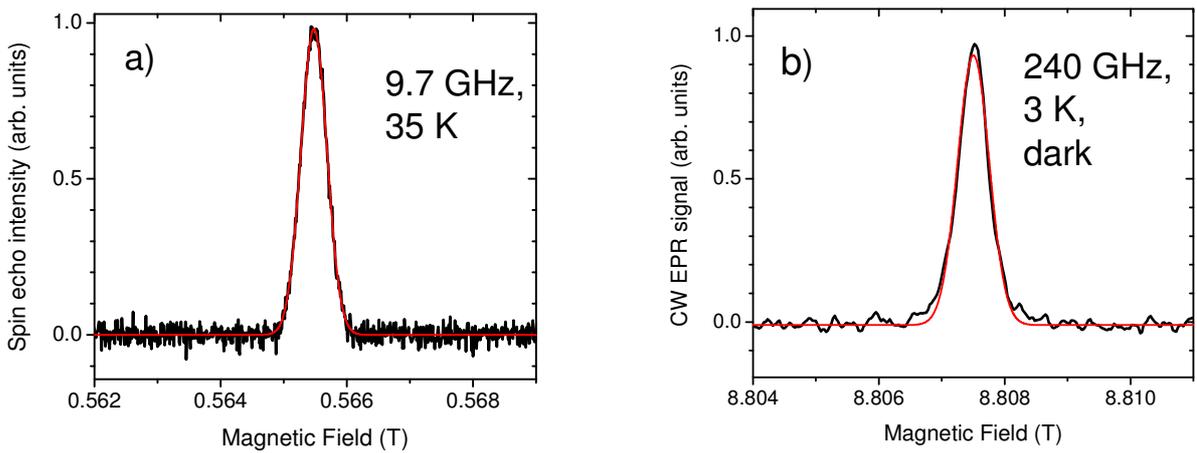

Figure S2. Expanded view of Figure 1 showing the highest-field resonance at a) 9.7 GHz and b) 240 GHz. The Gaussian fits have linewidths of 0.41 mT and 0.50 mT respectively due to unresolved hyperfine interactions with $^{29}$Si nuclei.

While resubmitting our manuscript for refereeing we found a preprint reporting similar experiments [supplementary ref 2].

## Supplementary References